\documentclass[conference]{IEEEtran}
\IEEEoverridecommandlockouts
\usepackage{cite}
\usepackage{amsmath,amssymb,amsfonts}
\usepackage{algorithmic}
\usepackage{graphicx}
\usepackage{textcomp}
\usepackage{xcolor}

\usepackage{dsfont}
\DeclareMathOperator*{\E}{\mathds{E}}

\DeclareMathOperator*{\eqdef}{\stackrel{\triangle}{=}}

\def\BibTeX{{\rm B\kern-.05em{\sc i\kern-.025em b}\kern-.08em
    T\kern-.1667em\lower.7ex\hbox{E}\kern-.125emX}}

\begin{document}

\title{A Proof of Concept Resource Management Scheme for Augmented Reality Applications in 5G Systems\\
\thanks{Identify applicable funding agency here. If none, delete this.}
}

\author{\IEEEauthorblockN{Panagiotis Nikolaidis\textsuperscript{1}, Samie Mostafavi\textsuperscript{2}, James Gross\textsuperscript{2} and John Baras\textsuperscript{1}}
\IEEEauthorblockA{\textit{\textsuperscript{1}Department of Electrical and Computer Engineering and the Institute for Systems Research, University of Maryland, USA}\\ \textit{\textsuperscript{2}School of Electrical Engineering and Computer Science, KTH Royal Institute of Technology, Sweden} \\
nikolaid@umd.edu, ssmos@kth.se, jamesgr@kth.se, baras@umd.edu}
}

\maketitle

\begin{abstract}
Augmented reality applications are bitrate intensive, delay-sensitive, and computationally demanding. To support them, mobile edge computing systems need to carefully manage both their networking and computing resources. To this end, we present a proof of concept resource management scheme that adapts the bandwidth at the base station and the GPU frequency at the edge to efficiently fulfill roundtrip delay constrains. Resource adaptation is performed using a Multi-Armed Bandit algorithm that accounts for the monotonic relationship between allocated resources and performance. We evaluate our scheme by experimentation on an OpenAirInterface 5G testbed where the considered application is OpenRTiST. The results indicate that our resource management scheme can substantially reduce both bandwidth usage and power consumption while delivering high quality of service. Overall, this work demonstrates that intelligent resource control can potentially establish systems that are not only more efficient but also more sustainable.
\end{abstract}


\section{Introduction}
Augmented Reality (AR) applications typically capture the user's natural environment with sensors and process it in real-time to add visual enhancements using Machine Learning (ML). Since these applications often run on mobile devices with limited processing power, their data is offloaded to a remote server for processing. 
Hence the AR data packets perform a roundtrip of three hops. First, they are sent to a remote server, second, they are processed at the server, and third, they are sent back to the user. Each hop contributes to the total roundtrip delay of the packet. However, the delay should remain low to provide a seamless user experience. 

To this end, Mobile Edge Computing (MEC) emerged \cite{MEC_surv, mobile_AR}. In MEC, the users send their data via 5G to a server close to the Base Station (BS), i.e., at the network's edge. MEC significantly reduces the networking delay compared to standard cloud computing. It also provides data locality, which is crucial for applications with sensitive user information.

Given that AR applications are bitrate-intense and delay-sensitive, delivering high Quality of Service (QoS) in MEC systems is a challenging multihop problem. It involves managing the system's networking and computing resources. The motivation for efficient resource management is two-fold. First, the resources are scarce. In particular, the licensed spectrum utilized by the BS is a costly resource that Mobile Network Operators (MNOs) obtain through competitive auctions. Moreover, computing resources are not as abundant as in cloud computing since they are deployed close to the user.

The second factor motivating efficient resource management is power consumption, which is affected by the number of resources used. For mobile devices, power consumption is crucial given their limited battery life. On the network side, the power consumed for data transmission and processing affects the Operating Expenses (OPEX) of the MNO and negatively impacts the environment. Hence efficient resource management is beneficial even when the traffic load is low and no resource contention arises. 

For this reason, we develop a multihop resource management scheme for AR running on MEC systems to utilize only the minimum amount of resources needed to meet the QoS requirements of all users. Specifically, we adapt online the allocated Physical Resource Blocks (PRBs) at the BS in uplink (UL) and downlink (DL), and the GPU frequency at the edge to the system's traffic load. We also consider a roundtrip delay budget to ensure that high QoS is delivered to the users.


The scheme performs mainly two operations. First, the roundtrip delay budget is split into three hop delay budgets. Second, at each hop, an online learning algorithm receives its hop delay budget and learns how to efficiently meet it by adapting the hop's resources. Here, we provide a static splitting of the roundtrip delay budget and we use Multi-Armed Bandits (MABs) to adapt the resources at each hop.

We implement our scheme on a testbed with Software Defined Radios (SDRs) that runs OpenAirInterface 5G, an open-source 5G software stack \cite{OAI}. The AR application that we consider is OpenRTiST which is also open-source \cite{openrtist}. We evaluate our scheme on 5 scenarios that are one hour long and compare it to three other baselines which results in more than 20 hours of experimentation. The results clearly show that our scheme supports fast resource adaption at the scale of a few seconds, provides high QoS, and significantly reduces the average allocated resources and power consumption at each hop. Lastly, we also provide detailed guides on Github to replicate our experimental setup as well as the network automation code that implements our proposed scheme \cite{installguide}.

The contributions of this paper are summarized as follows. To the best of our knowledge, we are the first to implement a multihop resource management scheme on an open-source MEC testbed that efficiently meets roundtrip delay constraints by adapting both networking and computing resources. In contrast to most related works, we do not use neural networks that must be pre-trained in reliable wireless network simulators since this approach is susceptible to the sim2real gap. Instead, we use a simple MAB algorithm. The algorithm learns a good policy fast by considering a monotonic relationship between the action and the cost since allocating more resources improves the QoS. Although this simple consideration greatly improves performance, it is omitted in most related works. 

\section{Related Literature}
A closely related research topic to this work is network slicing \cite{slice-survey}. Here, we essentially consider one slice of AR users and our goal is to allocate over time the minimum network and computing resources needed to meet a roundtrip delay constraint. Nonetheless, several works have investigated how the PRBs at the BS should be distributed among multiple slices when many customers are present \cite{channel-aware,application-aware,offlineRL}. Such PRB-sharing schemes are often presented as rApps or xApps \cite{cloudRIC, edgeRIC, collosseum} that run on the Radio Access Network Intelligent Controllers (RIC) within the Open Radio Access Network (O-RAN) architecture \cite{oran}. These PRB-sharing schemes typically involve ML methods for the control of network resources.

However, the previous works do not consider slicing MEC systems with computing resources for roundtrip delay constraints. Moreover, once the slicing problem becomes multihop, it is not clear how to extend them in a scalable manner. Due to these concerns, methods to decompose Service Level Agreements (SLA) that span multiple domains have been proposed in \cite{decomp1, decomp2}. Unfortunately, these methods have not been evaluated on real systems or network simulators hence several practical aspects such as the timescale of the decomposition algorithms are not fully addressed.

Here, we provide a static splitting method for the overall roundtrip delay budget that may be used to initialize dynamic SLA decomposition methods. Although we study only one slice in the MEC system, our approach can be extended to the case where multiple slices coexist. The extension is facilitated by the network slicing architecture in \cite{nsm1, nsm2} which proposes that the resource demand estimation task should occur separately for each slice. Moreover, the architecture resolves slice-level resource contention using only the fractions of the time that each slice has been denied service so far. All the above motivate the development of resource adaptation algorithms to occur for each slice separately.

On the computing side of MEC systems, several works consider offloading schemes such as \cite{poc-uav} and \cite{HI} that handle tradeoffs among power consumption, performance, and latency. Few works provide implementations of solutions that consider the network side of MEC systems. In \cite{HIMEC}, a radio-aware MEC system is implemented using OpenAirInterface LTE where the video-streaming quality is adapted based on channel conditions and congestion. Lastly, the authors in \cite{expeca} measure the traffic generated by OpenRTiST on a large and isolated testbed composed by both open-source components and proprietary hardware by Ericsson.  

Overall we were not able to find implementations of intelligent control schemes that manage both the networking and the computing resources on an open-source MEC testbed. To fill in this gap, we provide a proof-of-concept MEC system that can sense and adapt its resources based on the current traffic load in line with the vision for self-programming networks \cite{SPN}. The AR application and the resource management scheme presented here provide a tangible example of using ML for networks and networks for ML.

\section{The OpenRTiST Application}
\begin{figure}
\centering
\includegraphics[width=0.7\linewidth]{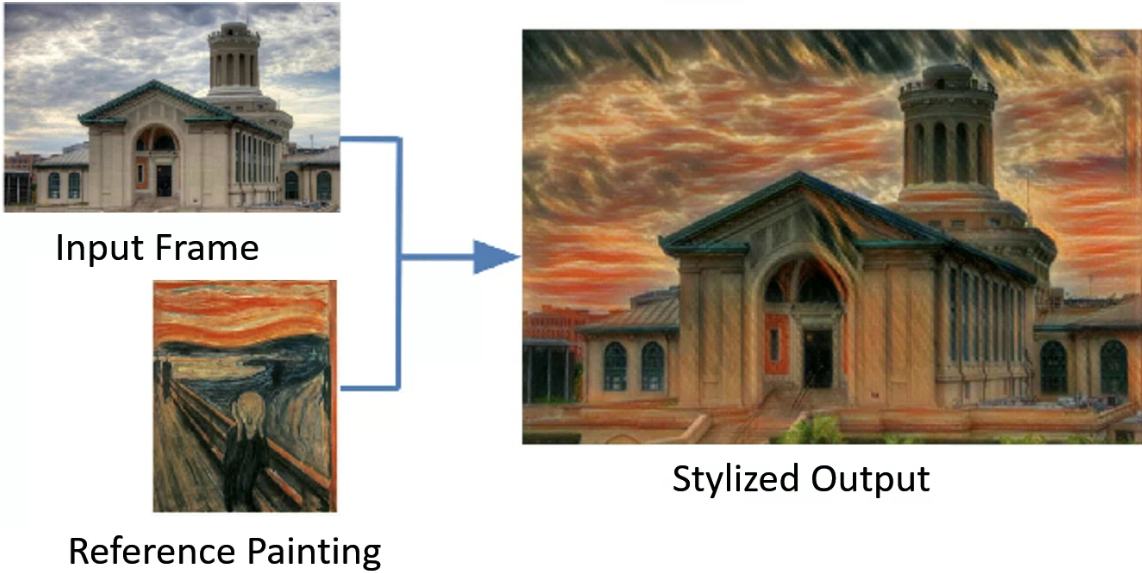}
\vspace{-1.8mm}
\caption{The OpenRTiST client at the user device sends frames to the server at the network's edge. The server then applies an artistic style to them and sends them back to the user. In our experiments, the client and server communicate via 5G. This figure is a modified version of Fig. 1 in \cite{openrtist}.}
\label{ortfig}
\end{figure}

We first describe the considered AR application, OpenRTiST \cite{openrtist}. The application is composed of an OpenRTiST client and an OpenRTiST server. The OpenRTiST client runs on the user’s device and receives video frames captured by the camera. Then, it sends these frames to the OpenRTiST server at the network's edge. The OpenRTiST server processes the frames and applies an artistic style to them. Lastly, the video frames are sent back to the OpenRTiST client at the user which completes the frame’s three-hop roundtrip. OpenRTiST's goal is to enable the user to "see the world through the eyes of an artist" as shown in Fig. \ref{ortfig}. 

As described in \cite{openrtist}, the processing of the frame involves a well-known Computer Vision (CV) technique called Neural Style Transfer (NST) \cite{nst1,nst2}. In short, OpenRTiST processes frames as follows. First, a Convolutional Neural Network (CNN) is trained offline to extract the style of a reference painting. Then, in real-time as the OpenRTiST server runs, inference on the CNN transforms the input frames. The pre-trained CNN used by OpenRTiST has 16 layers and performs approximately 2.4 billion operations when the input frame's resolution is $320 \times 240$.

As a result, OpenRTiST is a bitrate-intense, delay-sensitive, and computationally demanding application that motivates MEC setups.
Specifically, the application's traffic highly depends on the frame's resolution. The developers of OpenRTiST measured its traffic in UL and DL when the OpenRTiST client and server used a wireless protocol to communicate. Their results were presented in \cite[Table 2]{openrtist}. Table \ref{ort_og_measurements} summarizes some of their measurements.
\begin{table}
\centering
\caption{ OpenRTiST measurements by the developers \cite[Table 2]{openrtist}.}
\begin{tabular}{c c c c c}
\hline
Resolution & Frame Rate & UL Bitrate & DL Bitrate & Roundtrip\\
\hline
320 x 240 & 30 fps & 1.98 Mbps & 2.32 Mbps & 23 ms\\
480 x 360 & 30 fps & 2.97 Mbps & 4.22 Mbps & 44 ms\\
640 x 480 & 30 fps & 3.86 Mbps & 6.32 Mbps & 65.6 ms \\
960 x 720 & 30 fps & 4.36 Mbps & 8.56 Mbps & 165 ms
\end{tabular}
\label{ort_og_measurements}
\end{table}

As expected, the bitrates increase as the resolution increases. We note that OpenRTiST uses tokens to avoid network congestion. When one token is used, the next frame is sent to the server only once the previous frame is received. Thus, measuring the roundtrip frame delay does not fully describe the QoS since the frame sent rate may drop due to congestion. The fact that the frame sent rate may drop also affects the UL and DL bitrate measurements. As a result, tokens may hide the true demand of the application.

Due to the above, we conducted our own experiments to measure the true bitrate demand of OpenRTiST by disabling tokens to always force a $30$ fps frame sent rate. Furthermore, in our experiments, the OpenRTiST client and server are connected with a $750$ Mbps Ethernet cable to ensure that no network congestion occurs. Our results are shown in Table \ref{ort_measurements}.
\begin{table}
\centering
\caption{Our OpenRTiST measurements with a wired connection.}
\begin{tabular}{c c c c c}
\hline
Resolution & Frame Rate & UL Bitrate & DL Bitrate & Inference\\
\hline
320 x 240 & 30 fps & 11.7 Mbps & 3.7 Mbps & 28 ms \\
480 x 360 & 30 fps & 23.3 Mbps & 4.8 Mbps & 56.7 ms\\
640 x 480 & 30 fps & 36 Mbps & 5.1 Mbps & 96.7 ms \\
960 x 720 & 30 fps & 56 Mbps & 13 Mbps & 215.9 ms \\
1920 x 1080 & 30 fps & 85 Mbps & 6 Mbps & 645 ms\\
1280 x 720 & 30 fps & 70 Mbps & 4.8 Mbps & 280 ms\\
720 x 576 & 60 fps & 70 Mbps & 5 Mbps & 131 ms\\
720 x 480 & 30 fps & 60 Mbps & 4 Mbps & 110 ms\\
160 x 80 & 30 fps & 4.3 Mbps & 1.5 Mbps & 9 ms
\end{tabular}
\label{ort_measurements}
\end{table}
Our measurements indicate that the UL bitrate is significantly higher compared to the measurements in Table \ref{ort_og_measurements}. We believe that congestion in the wireless network affected the measurements obtained by the OpenRTiST developers. Congestion combined with the usage of tokens led to lower bitrate measurements in Table \ref{ort_og_measurements}.

Another interesting observation regarding Table \ref{ort_measurements} is that the DL bitrate is smaller than the UL bitrate even when there is no bottleneck in the network or compute side as the last entry of Table \ref{ort_measurements} shows. By inspecting the code of the OpenRTiST server, we noticed that the server compresses the stylized frames before sending them back to the user. However, compression does not happen before sending the original frames to the server. This is a possible explanation for the lower bitrate values in DL. Overall, the measurements show that OpenRTiST produces high traffic and compute loads.

\section{The Testbed}
\begin{figure}
\centering
\includegraphics[width=1\linewidth]{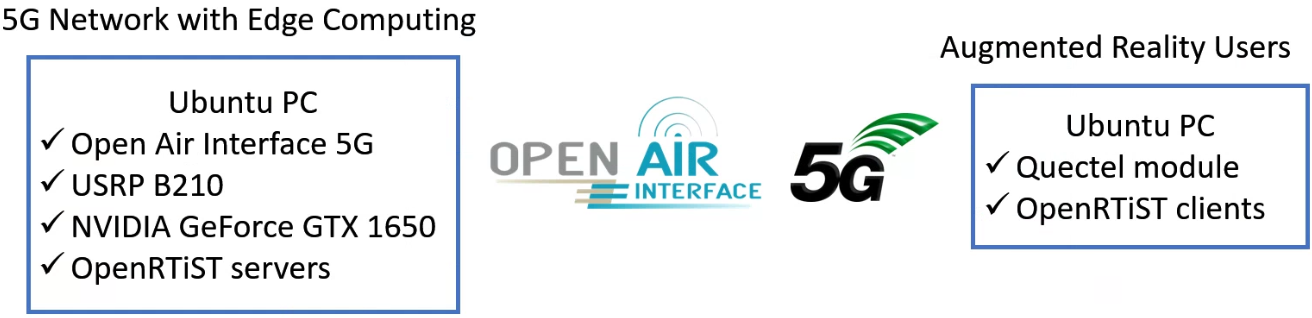}
\vspace{-7mm}
\caption{A diagram of our open-source testbed. The clocks of the two Ubuntu PCs are synchronized by PTP to accurately measure the delays incurred in UL, DL, and at the edge.}
\label{oaitestbed}
\end{figure}

 The 5G MEC testbed includes a USRP B210 SDR that is connected to an Ubuntu PC where OpenAirInterface 5G \cite{OAI} is executed to implement the 5G BS and Core Network (CN). The Ubuntu PC also hosts the OpenRTiST servers that process the frames using an NVIDIA GeForce GTX 1650 GPU. Hence, the edge and the 5G BS are co-located in our setup. To generate OpenRTiST traffic, we use another Ubuntu PC connected to a Quectel RM500Q-GL 5G HAT to implement a 5G User Equipment (UE). To accurately measure the delays incurred in UL, at the edge, and in DL, we synchronize the clocks of the two Ubuntu PCs using the Precision Time Protocol (PTP). An Intel Next Unit of Computing (NUC) hosts the PTP server. The testbed is depicted in Fig. \ref{oaitestbed}.

\begin{figure}
\centering
\includegraphics[width=0.9\linewidth]{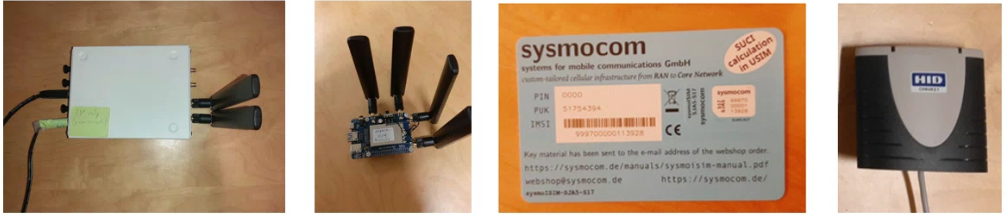}
\caption{From left to right: the USRP B210 SDR, the Quectel RF module, the SIM cards inserted at the Quectel module, and the SIM card reader to configure the cards. Our guide in \cite{installguide} describes how to setup the testbed in Fig. \ref{oaitestbed} using this hardware and Ubuntu PCs.}
\label{testbed-hardware}
\end{figure}

The steps needed to set up the testbed can be summarized as follows. First, a custom version of OpenAirInterface 5G needs to be installed to implement the 5G BS and CN. Our customized version includes network automation code that allows the dynamic adaptation of the PRBs used for DL and UL transmissions. Moreover, it enables the periodic extraction of network state quantities such as the number of active users, their channel conditions, and the number of queued bits. Second, the Quectel module in Fig. \ref{oaitestbed} needs to be configured via Attention (AT) commands and registered on the OpenAirInterface 5G database by configuring its Subscriber Identity Module (SIM). Third, modified versions of the OpenRTiST client and server must be installed on both PCs. Our modified OpenRTiST code timestamps the frames as they are sent by the user, received at the server, processed by the server, and finally received by the user. These four timestamps suffice to derive the delays incurred at each hop in the roundtrip journey of the frame and can be leveraged to determine where bottlenecks occur in the MEC system. Fourth, traffic generation code must be installed to start and stop OpenRTiST flows over long periods automatically. In Fig. \ref{testbed-hardware}, we depict some of the hardware needed to set up our testbed. A detailed setup guide can be found in \cite{installguide}. 

Next, we measure the capacity of the our 5G system to determine how many OpenRTiST flows can be supported. The system transmits on Band 78 and uses 106 PRBs with 30 KHz Sub Carrier Spacing (SCS). This results in 40 MHz of total utilized bandwidth. We do not use Multiple Input Multiple Output (MIMO) since it is not supported by OpenAirInterface 5G when using the USRP B210 SDR. 

The system operates in Time Division Duplexing (TDD) mode. As a result, the system's capacity is greatly affected by the TDD pattern; the number of DL and UL 5G slots within a 5G frame. The 5G frame is always fixed to 10 ms. Given that the SCS is 30 KHz in our setup, then the 5G slot duration is $0.5$ ms, and thus there are $20$ 5G slots in each 5G frame. We perform iperf3 measurements to determine the peak UL and DL bitrate for various TDD patterns. The results are shown in Table \ref{TDD-bitrate}. By comparing Table \ref{ort_measurements} to Table \ref{TDD-bitrate}, it is evident that the 5G system can support several OpenRTiST flows only when the $\rm{DSUUU}$ TDD pattern is used and when the OpenRTiST clients generate $30$ fps with $160 \times 80$ resolution. As a result, we run our experiments using these configurations.

\begin{table}
\centering
\caption{The capacity of our 5G system for various TDD patterns. "D", "U" and "S" refer to DL, UL, and Shared 5G slot respectively. }
\begin{tabular}{c c c}
\hline
TDD Pattern & UL Bitrate & DL Bitrate\\
\hline
DDDDDDDSUU &  12.5 Mbps & 99 Mbps\\
DDSUU &  16 Mbps & 82 Mbps\\
DSUUU & 22 Mbps & 44 Mbps
\end{tabular}
\label{TDD-bitrate}
\end{table}

\section{Problem Statement}
\label{probstat}
To provide a seamless user experience, the roundtrip frame delay should remain small. Moreover, to reduce the provisioned bandwidth at the BS and the power consumption of the MEC system, it is beneficial to adapt the allocated resources at each hop periodically based on the current traffic load.

Hence, we consider a time-slotted resource management scheme. Within each slot $t$, we wish to bound a metric of the frame delays sent within each slot $t$ by a target constant $Q_c$. The metric may be a statistic of the frame delays such as the average or the $90^{th}$ percentile. Let $\mathcal{F}(t)$ be the set of active flows throughout slot $t$. Let
a roundtrip frame delay metric of each active flow $f$ denoted by $Q_f(t)$. The overall QoS metric of the OpenRTiST flows at each slot $t$ is:
\begin{equation}
Q(t) \triangleq \prod_{f \in \mathcal{F}(t)}\mathds{1}(Q_f(t) \leq  Q_c),
\label{qoscontraint}
\end{equation}
where $\mathds{1}(\cdot)$ is the indicator function that returns $1$ if the argument is true and $0$ otherwise. Thus, QoS metric $Q(t)$ is a binary quantity that indicates whether the desired QoS is delivered at slot $t$. Next, we define the QoS delivery ratio that the OpenRTiST traffic experiences over an infinitely long time horizon $T_H$ as follows:
\begin{equation}
\bar{Q} \triangleq  \lim_{T_H \to \infty} \frac{1}{T_H}\sum\limits_{t=1}^{T_H}Q(t).
\label{qosdelratio}
\end{equation}

The main objective of the resource management scheme is to deliver a high QoS delivery ratio. However, we also wish that the allocated resources at each hop remain small. For each slot $t$, let $A^u(t)$, $A^e(t)$, and $A^d(t)$ denote the allocated UL PRBs at the BS, the GPU frequency at the edge, and the allocated DL PRBs at the BS. Similarly, as before, we define the average allocated bandwidth at each $h$:
\begin{equation}
\bar{A}^h \triangleq  \lim_{T_H \to  \infty} \frac{1}{T_H}\sum\limits_{t=1}^{T_H}A^h(t).
\label{avg_resources}
\end{equation}

The motivation for minimizing the average PRB allocation in UL and DL via resource adaptation is to increase the multiplexing potential of the OpenRTiST traffic. If other types of traffic are present in the system, the overall provisioned bandwidth at the BS may then be reduced.

However, multiplexing gains do not motivate the adaptation of the GPU frequency at the edge. Instead, the motivation here is reduced power consumption. Specifically, a model is provided in \cite[Eq. 7]{GPUWattch} stating that the GPU's power consumption is an affine function of its frequency $A^e(t)$:
\begin{equation}
P^e(t) \triangleq kA^e(t) + P^e_{\rm{const}}.
\label{gpumodel}
\end{equation}
The first term corresponds to the dynamic part of the GPU's power consumption. The second term is the GPU's power consumption when it is idle. The constants $k$ and $P^e_{\rm{const}}$ are obtained by fitting power measurements to (\ref{gpumodel}).  We consider (\ref{gpumodel}) because we could not obtain real power measurements from the GPU of our testbed. For the same reason, we cannot find constants $k$ and $P^u_{\rm{sleep}}$. As a result, in what follows, we consider only the average GPU frequency $\bar{A}^e$ which suffices to compute percentage savings for the dynamic part of the GPU's power consumption.

Reducing the power consumption is also beneficial at the BS since it reduces the OPEX of the MNO and the system's carbon footprint. We consider that the average allocated DL bandwidth $A^d(t)$ affects the power consumption at the BS in DL as in \cite[Eq. 1]{5Gpower}:
\begin{equation}
P^d(t) \triangleq \left(145 + 135\frac{A^d(t)}{A^d_{\rm{max}}}\right)P^d_{\rm{sleep}},
\label{bsconsumption}
\end{equation}
where $P^d_{\rm{sleep}}$ is the power consumption of the BS on sleep mode and $A^d_{\rm{max}}$ is the maximum number of PRBs available at the BS. As mentioned earlier, quantity $P^d(t)$ is the power consumption at the BS only for DL. We omit the BS power consumption in UL. 

Given the limited battery life in mobile devices, power consumption is critical on the UE side. In UL, the utilized bandwidth $A^u(t)$ affects power consumption \cite[Table 21]{uepower}:
\begin{equation}
P^u(t) \triangleq \left(0.4 + 0.6\left(40 \frac{A^u(t)}{A^u_{\rm{max}}} - 20\right)/80\right)P^u_{\rm{sleep}},
\label{uemodel}
\end{equation}
where $P^u_{\rm{sleep}}$ is the UE's power consumption in sleep mode.

Models (\ref{bsconsumption}) and (\ref{uemodel}) hold if the carrier Bandwidth Part (BWP) can be changed at each slot $t$. However, this is not currently supported by OpenAirInterface 5G. Nonetheless (\ref{bsconsumption}) and (\ref{uemodel}) provide potential power savings for 5G systems that support fast BWP adaptation with fine granularity. For more information regarding BWPs, we refer to \cite{bwp}.

Overall, the performance metrics considered here are the QoS delivery ratio $\bar{Q}$, the average UL and DL bandwidth $\bar{A}^u$ and $\bar{A}^d$, the average GPU frequency $\bar{A}^e$, and the average power consumption at the BS and the UE denoted by $\bar{P}^d$ and $\bar{P}^u$. Resource management schemes must consider tradeoffs among these $6$ metrics and ideally provide a Pareto optimal solution.

\section{Solution Approach}
As mentioned in the previous section, the main objective is maintaining a statistic of the roundtrip frame delays below a target threshold $Q_c$. We adapt the allocated resources at each hop based on the current traffic load in the 5G MEC system for increased efficiency. Notice that this problem involves a complex multihop queueing system. Specifically in the UL and DL hops, the 5G BS utilizes complicated MAC scheduling policies to distribute the PRBs among the currently active flows that balance total instantaneous bitrate and fairness. Due to this complexity, many closed-form results known from Queueing Theory are not applicable. As a result, we consider a data-driven approach. At the beginning of each slot $t$, we observe the state of the 5G MEC system $X(t)$, apply actions $A(t) \eqdef \{A^h(t)\}_{h \in \{d,e,u\}}$, and at the end of each slot $t$ receive QoS feedback $Q(t)$. Based on the observed sequence of tuples $(X(t),A(t),Q(t))$, a control policy can be learned.

\subsection{Online Learning}
To learn a control policy, we may first collect a long sequence of tuples $(X(t),A(t),Q(t))$ from long simulations or digital twins. Using this data, we may then obtain a function $f(x,a)$ that returns for each state-action pair $(x,a)$, either the pair's $Q$ value if Reinforcement Learning (RL) is used or the pair's expected round reward if contextual MABs are used. In all cases, the round reward $r(t)$ should depend on the QoS feedback $Q(t)$ and on the allocated resources $A(t)$. In most ML algorithms, function $f$ is typically a neural network. Once this function is computed, the optimal policy can be obtained and deployed on the real system.

Methods involving neural networks are promising since once the neural network is trained offline, only its parameters are stored in memory, not the system's dynamics. Unfortunately, there are two severe issues. First and foremost, wireless communications are hard to model reliably. Thus, high-fidelity simulations or digital twins are not readily available. Due to the sim2real gap, the model obtained via offline training may perform poorly on the real system. Second, even if the sim2real gap is bridged somehow, neural networks are known to be bad at interpolating unseen state-action pairs.

Due to the above, we propose an online learning method with few state-action pairs where we wish to learn a good policy fast. This is particularly important in queueing systems where an extended exploration phase may lead to large backlogs and users disconnecting from the system.

\subsection{Hop Decomposition}
Given that the OpenRTiST frames perform a roundtrip of three hops, the action space $\mathcal{A}$ is the Cartesian product of the hop action spaces, i.e., $\mathcal{A} =\mathcal{A}^d \times \mathcal{A}^e \times \mathcal{A}^u$. Since the action at each hop is a scalar, the action space $\mathcal{A}$ is three-dimensional. Similarly, the state space of the 5G MEC system can be seen as the Cartesian product of the hop state spaces $\mathcal{X} =\mathcal{X}^d \times \mathcal{X}^e \times \mathcal{X}^u$. However, in this case, each hop state space may be multidimensional. 

As a result, our learning problem involves a high dimensional state-action space $\mathcal{X} \times \mathcal{A}$ which requires large memory capacity since we do not use function approximation methods that depend on reliable offline training. To reduce the state-action space, we decompose the overall learning problem into three hop problems by splitting the roundtrip delay budget into three hop delay budgets:
\begin{equation}
\sum\limits_{h \in \{u,e,d\}}Q^h_c = Q_c.
\label{hop-delay-budget}
\end{equation}
Next, the hop QoS metrics are defined as follows:
\begin{equation}
Q^h(t) \triangleq \prod_{f \in \mathcal{F}(t)}\mathds{1}_{Q^h_f(t) \leq  Q^h_c}, \quad \forall h \in \{d,e,u\}.
\label{hop-delay-metric}
\end{equation}
The goal at each hop is to provide a high QoS delivery ratio:
\begin{equation}
\bar{Q^h} \triangleq  \lim_{T_H \to \infty} \frac{1}{T_H}\sum\limits_{t=1}^{T_h}Q^h(t), \quad \forall h \in \{d,e,u\}.
\label{hop-QoS}
\end{equation}

\begin{figure}
\centering
\includegraphics[width=1\linewidth]{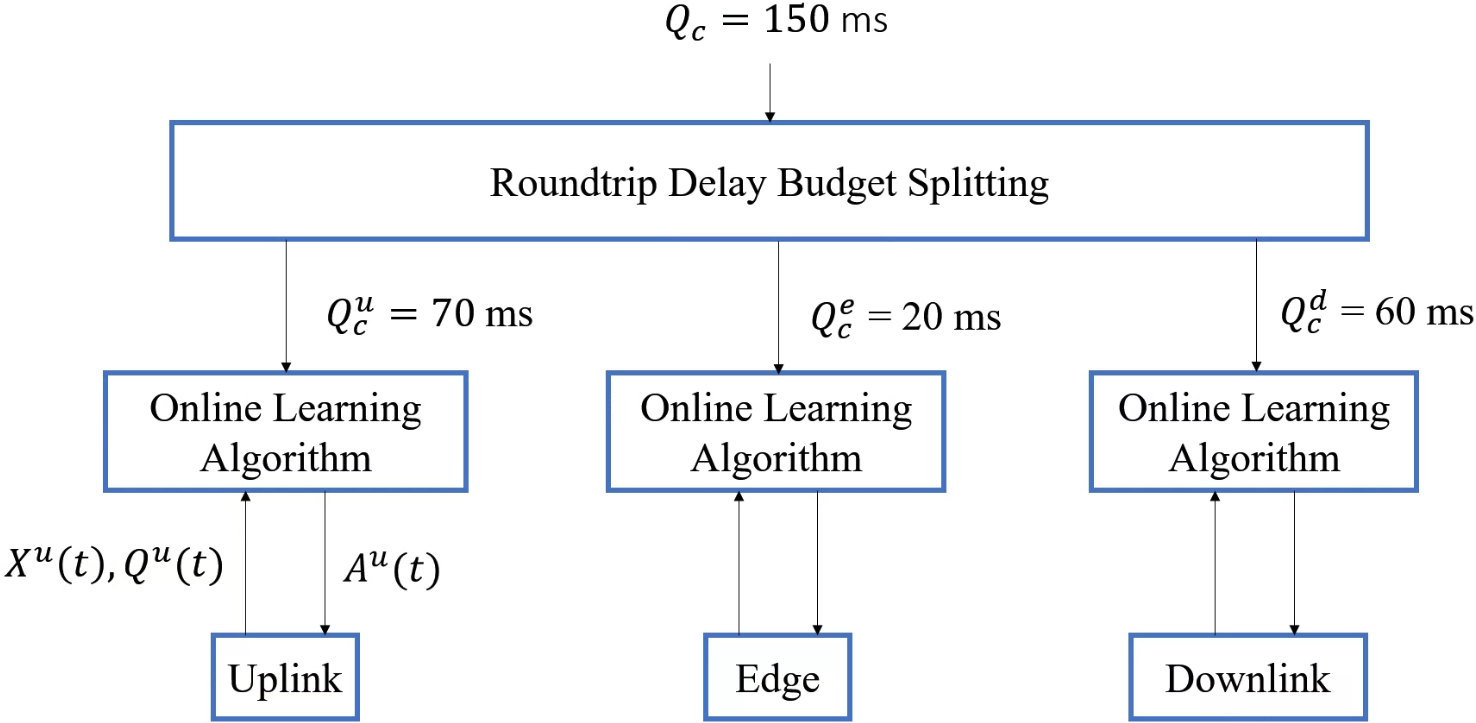}
\caption{We decompose the roundtrip learning problem into hop problems by splitting the roundtrip delay budget. Each hop employs an online learning algorithm to adhere to its delay budget efficiently.}
\label{per-hop-figure}
\end{figure}

Towards meeting the delay budgets $Q^h_c$, an online learning algorithm is deployed at each hop that efficiently learns how to adapt its resources $A^h(t)$ based on the hop's state $X^h(t)$ over time by periodically receiving hop feedback $Q^h(t)$. The architecture of the proposed solution is depicted in Fig. \ref{per-hop-figure}.

Although hop decomposition reduces the problem's complexity, notice that frames may satisfy the roundtrip delay budget but still violate a hop delay budget. Thus the hop decomposition strategy leads to performance loss. However, there is another motivating factor for hop decomposition. If each hop belongs to a different administrative domain, the hop's operator may only accept constraints involving their domain. Also, the operator may decline to share their domain's state $X^h(t)$ or permit direct control of their resources $A^h(t)$.

\section{Roundtrip Delay Budget Splitting}
Splitting the roundtrip delay budget into hop delay budgets is the first step in our solution approach. Notice that the smaller the delay budget is in a hop, the more hop resources must be allocated and the more leeway is provided in the other hops. Consequently, the MNO needs to rank the importance of each hop's resources and consider tradeoffs when splitting the budget. For instance, the MNO may consider a cost in dollars per resource unit for each hop. Then, an optimization problem can be solved to find the hop budgets that minimize the total cost subject to a high roundtrip QoS delivery ratio.

Unfortunately, methods to determine the dollar cost for each hop resource are hard to develop. For this reason, we consider a heuristic approach to split the delay budget. First, we run a simple experiment with only one OpenRTiST user using the maximum bandwidth in UL and DL, and the maximum GPU frequency at the edge. Then, we measure the delays incurred at each hop to consider the lowest possible hop delays that the 5G MEC system can provide. Table \ref{lowest_delays} presents the results.
\begin{table}
\centering
\caption{The frame delays at each hop when one OpenRTiST user is active and the allocated resources at each hop are maximized.}
\begin{tabular}{c c c c c}
\hline
Delay Metric & DL & Edge & UL & Roundtrip \\
\hline
Average & 19 ms & 9 ms & 12 ms & 40 ms\\
Minimum & 9 ms & 7 ms & 5 ms & 24 ms\\
Maximum & 62 ms & 376 ms & 216 ms & 567 ms  \\
$90^{th}$ percentile & 31 ms & 7 ms & 27 ms & 63 ms\\
\end{tabular}
\label{lowest_delays}
\end{table}

In case the QoS delay metric of interest is the average frame delay, Table \ref{lowest_delays} shows that if the desired bound is smaller than $Q_c=40$ ms, then the 5G MEC system cannot satisfy it for multiple OpenRTiST users. Here we consider that the desired QoS is to deliver an average frame delay less than $150$ ms.

We split the roundtrip delay budget $Q_c=150$ as follows. First, we consider that the UL bandwidth is the most costly resource since mobile devices are not as spectrally efficient as the BS and run on limited battery life. Next, the DL bandwidth is considered more costly than the computing resources since licensed spectrum is obtained through competitive auctions. Due to this, we consider that the ratio between the hop delay budget and the lowest possible hop delay should be as close to $5$, $3$, and $2$ for UL, DL and edge respectively. From Table \ref{lowest_delays}, it follows that $Q_c^u=70$ ms, $Q_c^d=60$ ms, and $Q_c^e=20$ ms as shown in Fig. \ref{per-hop-figure}.

For future work, we wish to develop an adaptive delay budget splitting method that operates at a slower time scale than the online learning algorithm in Fig. \ref{per-hop-figure}. The method should detect bottlenecks in case cross-traffic is present. Based on the traffic load at each hop, the method should adapt the delay budgets to provide more leeway to the bottleneck hop. Given that our testbed's capacity is too limited to study cross-traffic scenarios, we do not investigate adaptive budget splitting schemes here. Nonetheless, our static splitting method can be used to initialize adaptive methods based on single flow measurements as in Table \ref{lowest_delays}.

\section{The Online Learning Algorithm}
\label{propola}
Our solution centers on an online learning algorithm deployed at each hop that adapts the allocated hop resources to efficiently meet the hop delay constraint. The algorithm observes the hop state $X^h(t)$ at the start of slot $t$, allocates hop resources $A^h(t)$, and receives QoS feedback $Q^h(t)$ at the end of slot $t$ for the frame delays in slot $t$. 

We note that each hop involves a queueing system. In such cases, allocating resources $A^h(t)$ affects not only the current QoS $Q^h(t)$ but also future ones since the next state of the system $X^h(t+1)$ largely depends on the previous action $A^h(t)$. For instance, if low resources are allocated at slot $t$, large backlogs may be created in the queue at slot $t+1$. As a result, an RL approach is suitable for considering that the actions affect the system's trajectory. 

However, we propose a MAB algorithm, primarily for simplicity and ease of implementation. MAB algorithms are myopic since they greedily select the action with the lowest expected round cost without considering the future trajectory of the system. However, in queueing systems, we argue that this is not an unreasonable approach. Indeed, if an action at the end of the slot $t$ delivered low frame delays, then the queue length of the system is most likely small. This implies that MAB policies most likely transition the system to promising states. Hence myopic MAB solutions may achieve performance comparable to RL policies but with less complexity.

Due to the above, we propose Monotonic Upper Confidence Bound 1 (MUCB1), a UCB1 \cite{ucb1} variant that considers that allocating more resources improves the QoS. In what follows, we describe in detail how the state, action, and cost are defined in MUCB1 and how they are implemented on the testbed.

\subsection{State}
The state summarizes the system's history. It also specifies what the algorithm adapts to. Multiple factors affect the 5G MEC system such as user traffic, channel conditions, and queue lengths. For instance, we may use a vector for each active UE in the system that includes the user's SNR, velocity, bits in the MAC layer, and statistics for packet inter-arrival and service times. However, this results in a large multi-dimensional state whose dimensions depend on the maximum number of active users.

\subsubsection{Hop state in UL and DL}
To reduce the size of these hop states, we leverage the layered structure of 5G NR. Since bandwidth allocation is performed at the MAC layer, we do not consider Physical (PHY) layer quantities in our state. In the MAC layer, the channel conditions of a user are summarized by the MCS index which describes the bits per PRB that can be transmitted to achieve a certain Block Error Rate (BLER). The MCS index provides information regarding the spectral efficiency which can be used to compute the time to serve a certain amount of bits given a certain amount of allocated bandwidth in the queueing system.

However, it is also needed to describe the traffic load in the system hence we also monitor the incoming data bits in the system and the queued bits. Then, using the MCS indices of the active users, we compute the number of PRBs required to transmit the sum of the incoming bits and the queued bits. Hence, the hop states $X^u(t)$ and $X^d(t)$ are scalars measured in PRBs within interval $[0,106]$. The last step is to round up or down the values of $X^u(t)$ and $X^d(t)$ to multiples of $5$ to create a discrete state space similar to the action spaces $\mathcal{A}^u$ and $\mathcal{A}^d$ in UL and DL. The hop states approximate the bandwidth needed to match the incoming bitrate. 

More specifically, the hop states in UL and DL at the start of slot $t$ are computed as follows. Throughout the previous slot $t-1$ that lasts $5$ seconds, we consider $25$ sub-intervals of $200$ ms. For each sub-interval, we measure the incoming bits in UL and DL at the Radio Link Control (RLC), layer and the queued bits at the MAC layer for each active user along with their MCS index. Based on these measurements, we compute the number of PRBs needed to transmit all these bits within $200$ ms. Then, at the end of the $25^{th}$ sub-interval, we average all these $25$ numbers of PRBs. The two averages in UL and DL are the hop states at the start of slot $t$ in UL and DL. 

Upon experimentation, we observed that the bits arriving at the RLC layer do not correspond to the number of data bits generated by OpenRTiST. It is possible that the number of OpenRTiST data bits can only be measured at the Packet Data Convergence Protocol (PDCP) layer of 5G. However, we could not find such measurements at the PDCP layer in OpenAirInterface 5G. Due to the above, we did not use the RLC measurements mentioned in the previous paragraph. Instead, we measured the OpenRTiST bitrate offline and then used this measurement as the incoming bits within $200$ ms.

\subsubsection{Hop state at the edge}
The edge involves a system where incoming OpenRTiST frames await processing by the GPU. Unfortunately, we could not extract the number of queued frames at the GPU. Instead, we approximate the state of the GPU $X^e(t)$ by the state in UL $X^u(t)$, i.e., $X^e(t)=X^u(t)$. 

\subsection{Action}
As mentioned previously the action is the allocation of PRBs in UL and DL and the GPU frequency used at the edge. In all cases, the action space is discrete. Specifically, we consider $\mathcal{A}^d = \mathcal{A}^u = \{10,20,30...,100, 106\}$ PRBs and $\mathcal{A}^e = \{500,600,...,1600\}$ MHz.

\subsection{Cost}
The $c^h(t)$ at each slot $t$ and hop $h$ involves both the binary QoS metric $Q^h(t)$ of (\ref{hop-delay-metric}) and the allocated resources $A^h(t)$:
\begin{equation}
c^h(t) = A^h(t) + \lambda^h (1-Q^h(t)),
\label{slot-cost}
\end{equation}
where $\lambda^h$ handles the tradeoff between average allocated resources and QoS delivery, two of the metrics mentioned in Sec. \ref{probstat}. Given that $Q^h(t)$ is binary, notice that $c^h(t)=A^h(t)$ if the QoS was delivered. Otherwise $c^h(t)=A^h(t) + \lambda^h$. To determine the value of $\lambda^h$, notice that the optimal strategy eventually learned by the MAB algorithm is to select the action with the smallest expected cost given the current state-action pair. Using (\ref{slot-cost}), it follows:
\begin{equation}
c^h_{x,a} \triangleq \E[c^h(t)|X^h(t)=x, \, A^h(t)=a] =  a + \lambda^h Q^h(x,a),
\label{expected-slot-cost}
\end{equation}
where $Q^h(x,a)$ is the delay violation probability when the state-action pair is $(x,a)$, i.e., the probability that the hop delay exceeds the hop delay budget given that the hop state is $x$ and the action is $a$. Thus, we may make an action $a$ more appealing at state $x$ by configuring $\lambda^h$ to lower the expected cost $c^h_{x,a}$. Here, we choose $\lambda^h = 0.2A^h_{\max}/0.1$ which implies that we are willing to trade $20\%$ of the total resources of hop $h$ if the hop delay violation probability decreases by $10\%$.

\subsection{Monotonicity}
As mentioned earlier, the proposed algorithm MUCB1 is a variant of the standard UCB1 algorithm in \cite{ucb1} that considers the monotonicity between allocated resources and QoS feedback. Specifically, we consider that if $A^h(t)$ failed the hop delay constraint and resulted in cost $c^h(t)=A^h(t)+\lambda^h$, then all smaller actions $a \leq A^h(t)$ would also fail it and would have costed $a + \lambda^h$. Similarly, if $A^h(t)$ satisfied the hop delay constraint and achieved cost $A^h(t)$, then all larger actions $a \geq A^h(t)$ would also satisfy it and achieve cost $a$. Hence, unlike the standard UCB1 algorithm, at each slot $t$ we received feedback for multiple actions not just for the selected action $A^h(t)$. This has a significant impact on performance since it reduces exploration. We note that the exploitation of monotonicity in the context of bandwidth allocation was first proposed in a MAB algorithm in \cite{bde}. However, the previous algorithm used a three-dimensional state which increased the computational complexity. Lastly, the network automation code implementing our solution approach is available on Github\footnote{https://github.com/pnikolaid/network-control/blob/main/README.md}.

\section{Performance Evaluation}

\subsection{Metrics}

\textbf{QoS Delivery Ratio}: The goal of our scheme is to ensure that the average roundtrip frame delay is below $Q_c=150$ ms for a high fraction of slots. The QoS delivery ratio $\bar{Q}$ is computed using (\ref{qosdelratio}).

\textbf{Average UL PRBs}: The scheme should efficiently adapt the bandwidth in UL to achieve high resource utilization and multiplexing gains in case other types of traffic are present in the 5G network. The average allocated bandwidth in UL $\bar{A}^u$ is computed as in (\ref{avg_resources}).

\textbf{Average DL PRBs}: We wish to minimize the DL bandwidth for multiplexing gains. The average allocated bandwidth in DL $\bar{A}^d$ is computed as in (\ref{avg_resources}).

\textbf{Average GPU Frequency}: The reduction of the GPU frequency results in power savings at the edge as follows from (\ref{gpumodel}). We depict directly the average GPU frequency $\bar{A}^e$ as in (\ref{avg_resources}) which can used to find percentage savings for the dynamic power of the GPU.

\textbf{Average BS Power Savings}: The power consumption at the BS is computed using (\ref{bsconsumption}) and it is an affine function of the DL bandwidth $A^d(t)$. Power savings are calculated compared to the scenario where the DL bandwidth is consistently at its maximum, i.e., $A^d(t) = 106$ PRBs.

\textbf{Average UE Power Savings}: The power consumption at the UE is computed using (\ref{uemodel}). Similarly, it is an affine function of the UL bandwidth $A^u(t)$. Power savings are calculated compared to the scenario where the UL bandwidth is always maximum, i.e., $A^u(t) = 106$ PRBs.

\subsection{Schemes}
We compare the performance of our scheme by considering several baselines for the Online Learning Algorithm deployed at each hop. We consider the following four schemes.

\textbf{Static:} This scheme constantly allocates the maximum amount of resources at each hop, i.e., $A^u(t)=A^d(t)=106$ PRBs and $A^e(t)=1600$ MHz $\forall t$. The static scheme provides the highest possible QoS delivery ratio.

\textbf{TCP-based:} It is not always necessary to allocate resources using online learning algorithms. A tried and true method in networking is the Additive Increase and Multiplicative Decrease (AIMD) used by the Transmission Control Protocol (TCP). In short, when packet drops occur, TCP drastically reduces the sender's congestion window by dividing it by a constant. In case of no packet drops, TCP slowly increases the congestion window by a constant. Similarly, we consider a TCP-based scheme that doubles the allocated resources when the QoS is violated, i.e., when $Q(t)=0$, and reduces the resources to the next smallest value when the QoS is satisfied, i.e., when $Q(t)=1$. The TCP-based scheme provides a simple baseline that does not utilize online learning.

\textbf{UCB1:} This scheme utilizes the standard UCB1 algorithm with the state, action and cost as defined in Sec. \ref{propola}. It provides an online learning baseline.

\textbf{MUCB1:} This is the proposed scheme of Sec. \ref{propola} that considers that more resources achieve better performance. Comparisons with the UCB1 scheme reveal the importance of incorporating monotonicity in the algorithm.

\subsection{Scenarios}
We compare the schemes across all six metrics in $5$ different test scenarios. We create time-varying traffic with either $2$ or $3$ OpenRTiST users whose on and off durations follow exponential distributions. Our testbed cannot support more users. Each scenario lasts more than $1$ hour resulting in total experimentation time of more than $20$ hours. The slot length of the adaptation algorithm is in all cases $5$ seconds. Hence the resources are adapted at least $720$ times in each scenario. This results in approximately $240$ rounds per traffic load. For each scenario, we provide a figure detailing the performance of the MUCB1 scheme and then we present how it performs against the other three baselines for all the six metrics in Table \ref{king_table}. The parameter files for the scenarios can be found on Github\footnote{github.com/pnikolaid/network-control/blob/main/scenario\_parameters}.

\begin{figure}
\centering
\includegraphics[width=0.95\linewidth]{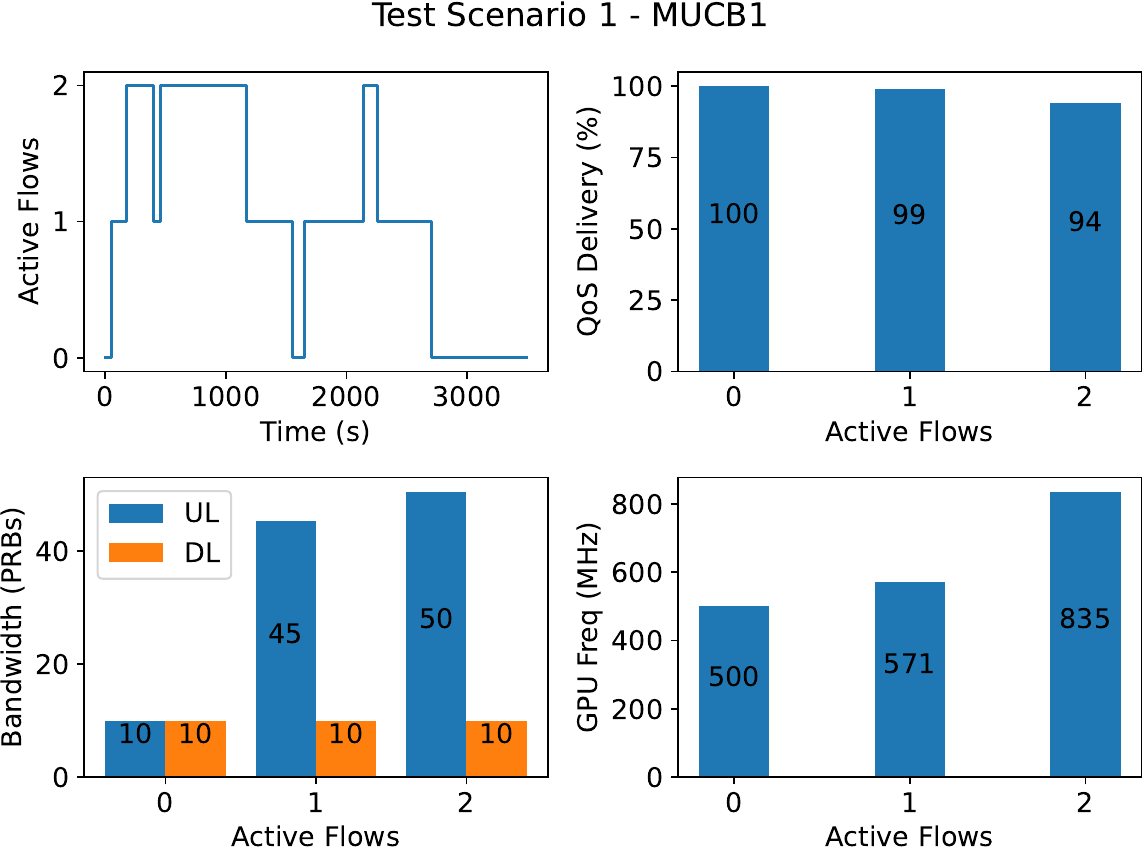}
\vspace{-8pt}
\caption{For each traffic load, we depict the last $100$ allocations made by the MUCB1 to show what the algorithm eventually learned.}
\label{ts1_MUCB1}
\end{figure}

\textbf{Test Scenario 1:} We consider two OpenRTiST users. Each user has on and off periods with means $5$ and $4$ minutes respectively. We also set min on and off durations of $2$ minutes. The scenario's duration is $1$ hour. The upper left plot in Fig. \ref{ts1_MUCB1} shows how the number of active flows varies over time. The other three plots show the performance of MUCB1 for the various traffic loads. For each load, we show an average quantity of the most recent $100$ allocations to show where the MUCB1 converged. The upper-right plot shows that MUCB1 provides a high QoS delivery ratio for all loads. Similarly, the bottom two plots show that MUCB1 learns to adapt the bandwidth. As we expected, the higher the traffic load the more the allocated resources. In DL, $10$ PRBs suffice for all traffic loads since the DL rate is just $1.5$ Mbps for $160 \times 80$ frame resolution in Table \ref{ort_measurements}. Next, Table \ref{king_table} contains comparisons that consider all the experiment's data. It shows that MUCB1 outperforms all adaptation baselines in both QoS delivery ratio and resource efficiency. Also, it provides a high QoS delivery ratio of $93\%$ which is only $2\%$ lower than that of the Static scheme that provides the highest possible QoS.

\begin{figure}
\centering
\includegraphics[width=0.95\linewidth]{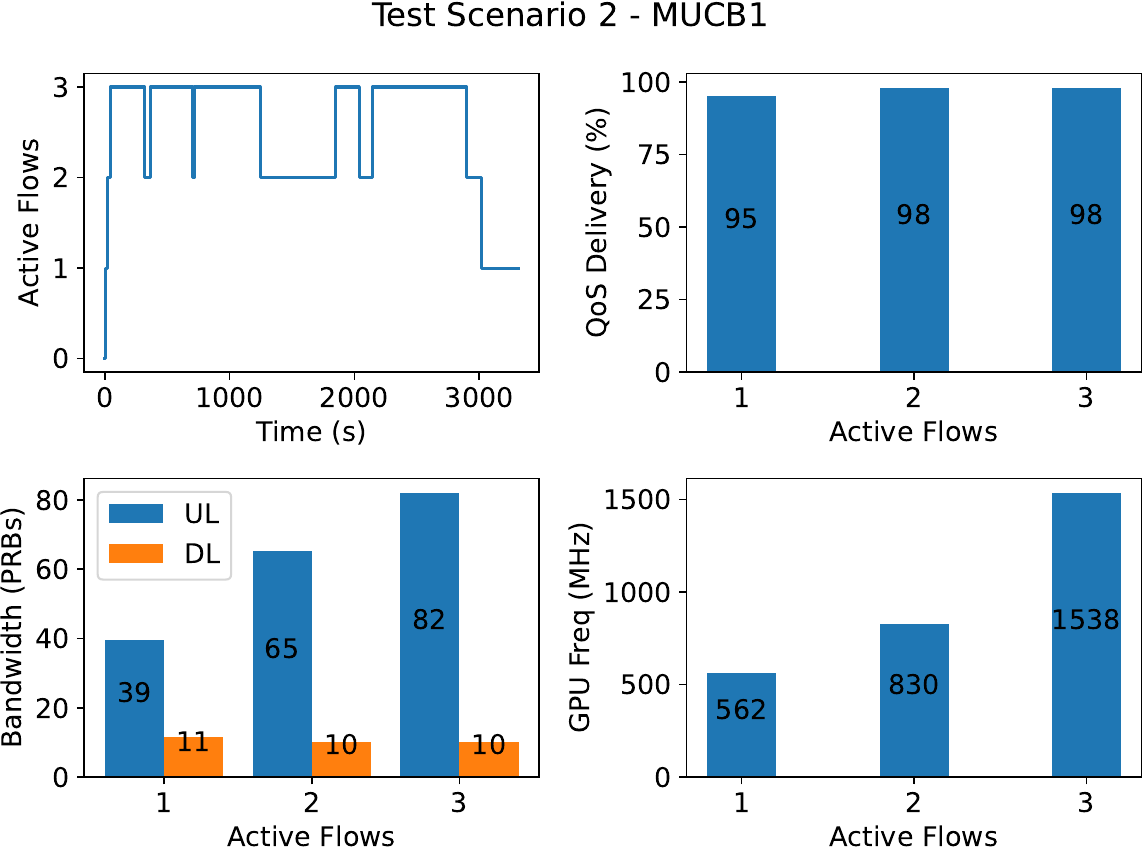}
 \vspace{-8pt}
\caption{We do not depict the case of no active flows in the system since this happens only for a few seconds as the upper left plot shows. }
\label{ts2_MUCB1}
\end{figure}

\textbf{Test Scenario 2:} Here we consider $3$ OpenRTiST users with mean on and off periods of $10$ and $1$ minutes respectively. The minimum on and off durations are $5$ and $1$ minutes. The scenario length is $1$ hour. The UL and DL action space is $\{10, 15, 20,.. 105\}$. Figure \ref{ts2_MUCB1} shows that MUCB1 efficiently reduces the allocated resources while maintaining high QoS. Table \ref{king_table} shows that overall the QoS delivery ratio of the MUCB1 is $88\%$. However, if only the last $100$ values are considered per traffic load as in Fig. \ref{ts2_MUCB1}, the QoS is above $95\%$.

\textbf{Test Scenario 3:} The scenario lasts $1$ hour and includes $2$ users with mean on and off times of $5$ and $2$ minutes respectively. However, the UL and DL action space is $\{10, 20, 40, 60, 80, 100\}.$ Figure \ref{ts1_MUCB1} shows that the MUCB1 algorithm achieves high QoS efficiently. Table \ref{king_table} shows that MUCB1 provides higher QoS than all baseline schemes.
\begin{figure}
\centering
\includegraphics[width=0.95\linewidth]{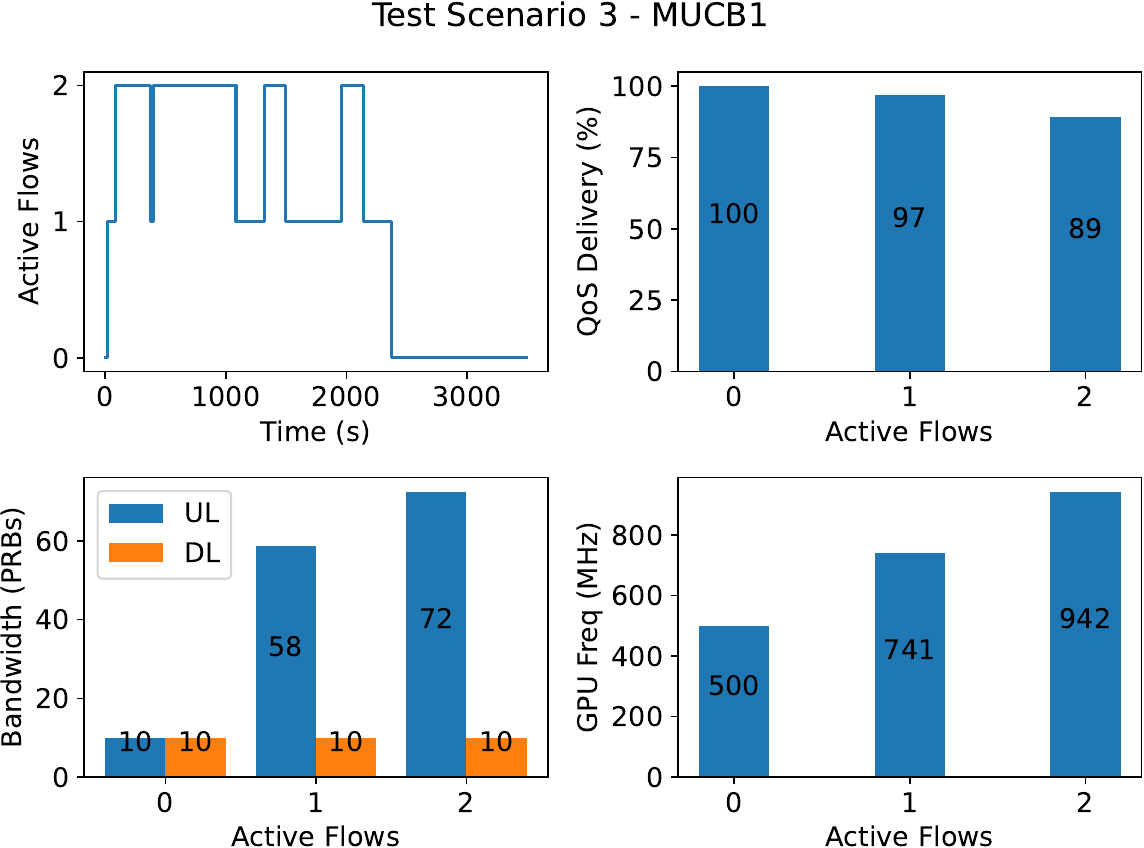}
 \vspace{-8pt}
\caption{The MUCB1 algorithm learns to adapt the resources efficiently. As expected, a higher traffic load results in more allocated resources.}
\label{ts3_MUCB1}
\end{figure}

\begin{figure}
\centering
\includegraphics[width=0.95\linewidth]{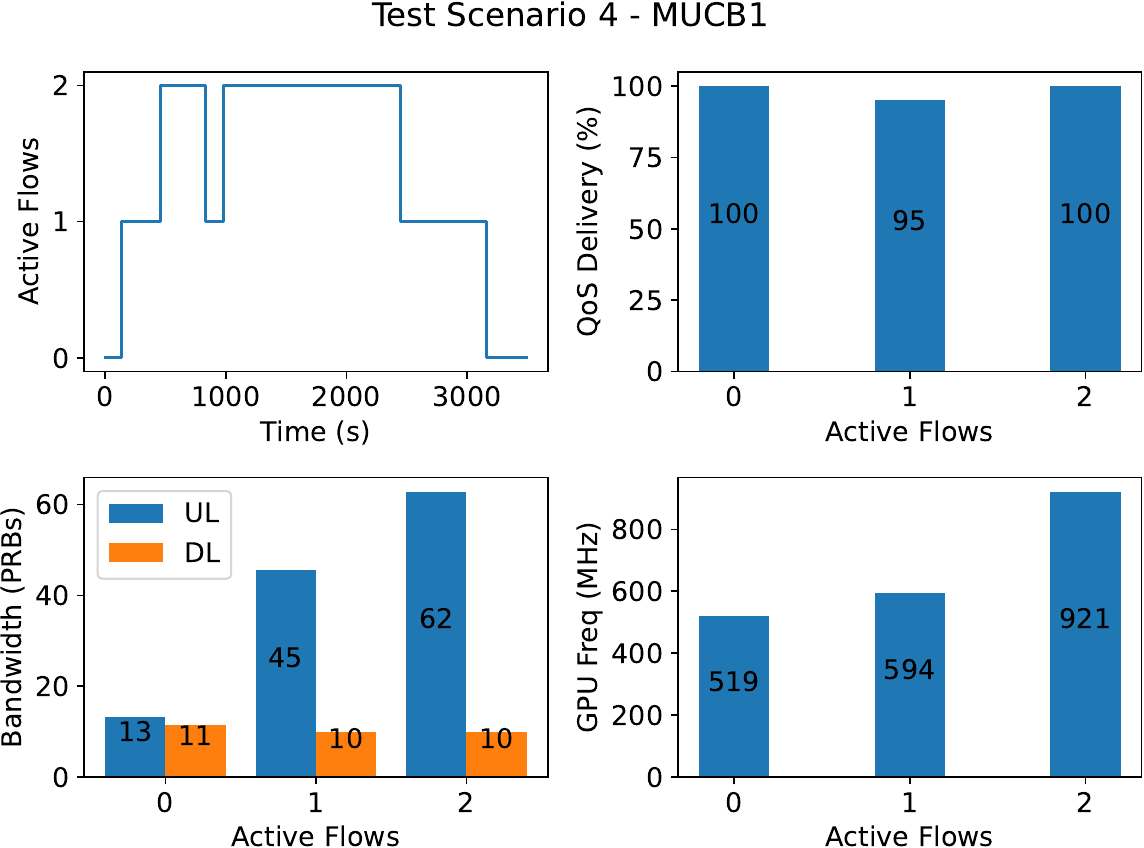}
 \vspace{-8pt}
\caption{For two active flows, the QoS delivery ratio is $100\%$ possibly because the last $100$ times the MUCB1 algorithm picked a high enough bandwidth.}
\label{ts4_MUCB1}
\end{figure}

\textbf{Test Scenario 4:} Two OpenRTiST users are considered with mean on and off periods of $10$ minutes. The action space in UL and DL is $\{10, 20, 30, ..., 100\}$ PRBs. Figure \ref{ts4_MUCB1} depicts the performance of MUCB1 and Table \ref{king_table} shows that MUCB1 delivered a QoS delivery ratio of $93.3\%$. 

\begin{figure}
\centering
\includegraphics[width=0.95\linewidth]{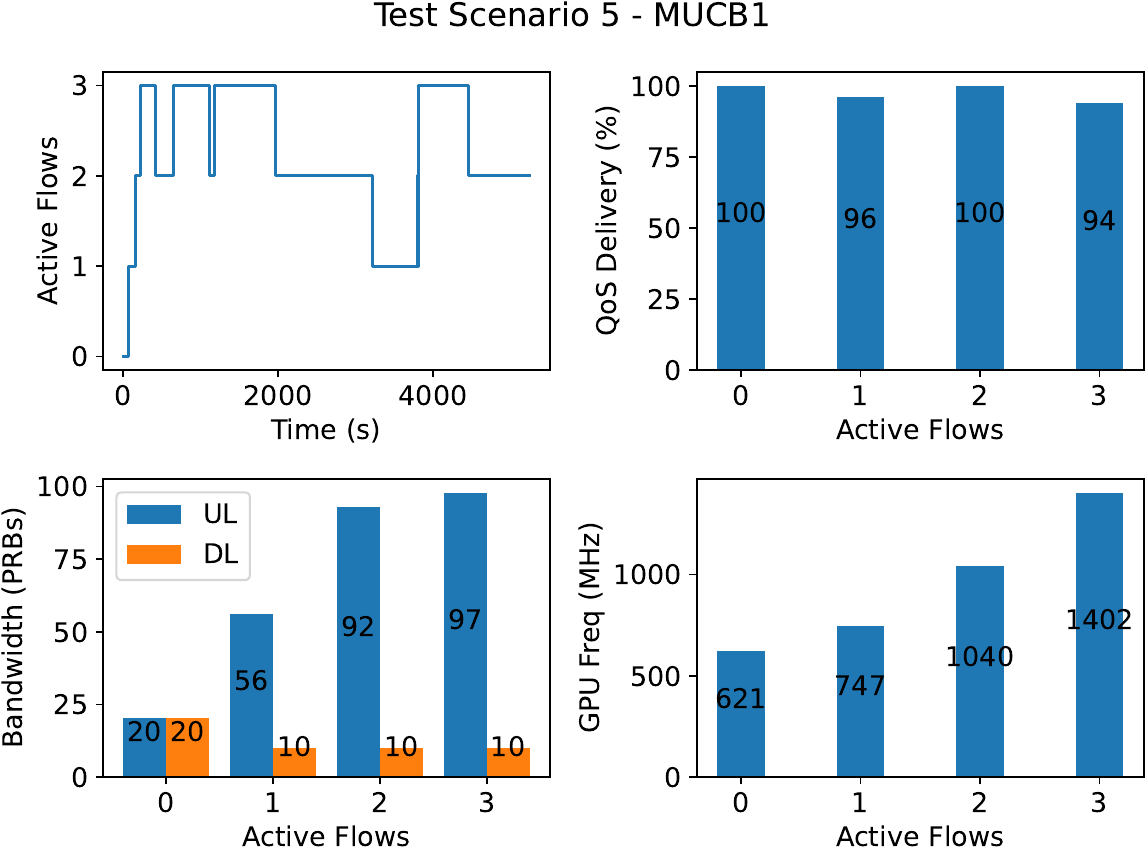}
 \vspace{-8pt}
\caption{The high QoS delivery ratios of the upper right plot show that MUCB1 eventually learns how to adapt the resources.}
\label{ts5_MUCB1}
\end{figure}

\textbf{Test Scenario 5:}
We conclude our evaluation with a scenario that lasts $1.5$ hours with $3$ OpenRTiST users. The action space in UL and DL $\{10, 20, 30, ..., 100\}$. Figure \ref{ts5_MUCB1} shows that MUCB1 eventually achieves more than $90\%$ delivery for each traffic load even though Table \ref{king_table} shows $85\%$. As mentioned earlier, Table \ref{king_table} includes the initial exploration phase of the algorithm that negatively impacts QoS.

\begin{table}
\centering
\caption{The results consider the whole system trajectory. The MUCB1 QoS delivery ratio is below $90\%$ in scenarios 3 and 5. However, Fig. \ref{ts3_MUCB1} and \ref{ts5_MUCB1} show that MUCB1 eventually provides high QoS.}
\begin{tabular}{c  c  c  c  c  c  c}
\hline
Scheme & QoS & \multicolumn{2}{c}{BW (PRBs)} &  Freq (MHz) & \multicolumn{2}{c}{Power Savings}\\
&& UL & DL & & UE & BS\\
\hline
\\
\multicolumn{7}{c}{\textbf{Test Scenario 1}} \\
\hline
Static & 95.0\% & 106 & 106 & 1600  & 0\% & 0\% \\
TCP-based & 84.1\% & 27 & 10 & 809  & 40\% & 43\% \\
UCB1 & 86.0\% & 45 & 36 & 1018  & 30\% & 31\% \\
MUCB1 & 92.9\% & 33 & 10 & 623  & 37\% & 43\% \\
\\
\multicolumn{7}{c}{\textbf{Test Scenario 2}} \\
\hline
Static & 96.9\% & 106 & 106 & 1600  & 0\% & 0\% \\
TCP-based & 90.0\% & 75 & 13 & 1122  & 15\% & 42\% \\
UCB1 & 41.8\% & 57 & 50 & 1076  & 25\% & 25\% \\
MUCB1 & 87.8\% & 70 & 11 & 1177  & 18\% & 43\% \\
\\
\multicolumn{7}{c}{\textbf{Test Scenario 3}} \\
\hline
Static & 97.0\% & 106 & 106 & 1600  & 0\% & 0\% \\
TCP-based & 79.6\% & 30 & 11 & 912  & 38\% & 43\% \\
UCB1 & 85.4\% & 46 & 35 & 1019  & 30\% & 31\% \\
MUCB1 & 92.9\% & 42 & 10 & 716  & 32\% & 43\% \\
\\
\multicolumn{7}{c}{\textbf{Test Scenario 4}} \\
\hline
Static & 95.9\% & 106 & 106 & 1600  & 0\% & 0\% \\
TCP-based & 88.0\% & 47 & 10 & 866  & 30\% & 43\% \\
UCB1 & 86.0\% & 54 & 45 & 1043  & 26\% & 27\% \\
MUCB1 & 93.3\% & 48 & 10 & 725  & 29\% & 43\% \\
\\
\multicolumn{7}{c}{\textbf{Test Scenario 5}} \\
\hline
Static & 96.2\% & 106 & 106 & 1600  & 0\% & 0\% \\
TCP-based & 86.8\% & 65 & 10 & 1162  & 20\% & 43\% \\
UCB1 & 58.7\% & 54 & 42 & 1065  & 26\% & 28\% \\
MUCB1 & 85.1\% & 78 & 10 & 1061  & 14\% & 43\% \\
\end{tabular}
\label{king_table}
\end{table}

\subsection{Summary of Results}
As expected, the scenario plots verify that higher traffic loads require more resources. They also also show that MUCB1 delivers high QoS during low and high traffic loads. The DL bandwidth is on average $10$ PRBs across all scenarios since the OpenRTiST DL bitrate is only 1.5 Mbps as shown in Table \ref{ort_measurements}.  Table \ref{king_table} shows that MUCB1 greatly outperforms the standard UCB1 algorithm which highlights the importance of exploiting monotonicity to achieve good performance fast.  The TCP-based scheme sligthly outperforms the MUCB1 scheme in scenarios $2$ and $5$. However, figures \ref{ts3_MUCB1} and \ref{ts5_MUCB1} show that the MUCB1 eventually achieves high QoS delivery ratios. Thus, the MNOs may need to tolerate subpar performance from the online learning schemes for a short period of time since these schemes must explore the action space. In general, it is known that the expected regret of the standard UCB1 algorithm scales logarithmically over time and depends on the number of actions. Here, the results show that within $250$ rounds per state, the MUCB1 algorithm achieves high QoS. Lastly, Table \ref{king_table} shows that the MUCB1 algorithm achieves a QoS delivery ratio close to the one achieved by the static scheme but with significantly less resources.

\section{Conclusion}
The strict delay requirements of AR applications combined with the resource scarcity in MEC systems calls for efficient multihop resource control. We presented a proof of concept autonomous MEC system that efficiently adapts its resources to satisfy roundtrip delay constraints. Unlike most related works, our approach does not rely on neural networks that require reliable network simulators for training. Instead, we developed a simple MAB algorithm that learns a good policy fast by leveraging the fact that more resources lead to better QoS. Experimental results showed that our approach provides an appealing tradeoff among the QoS delivery ratio, allocated resources and power consumption. For future work, we wish to consider dynamic hop delay budgets based on hop congestion.

\bibliographystyle{IEEEtran}
\bibliography{references}

\end{document}